# A novel optical assay system for bilirubin concentration measurement in whole blood

Jean Pierre Ndabakuranye, Anushi E. Rajapaksa, Genia Burchall, Shiqiang Li, Steven Prawer, Arman Ahnood

*Abstract*— As a biomarker for liver disease, bilirubin has been utilized in prognostic scoring systems for cirrhosis. While laboratory-based methods are used to determine bilirubin levels in clinical settings, they do not readily lend themselves to applications outside of hospitals. Consequently, bilirubin monitoring for cirrhotic patients is often performed only intermittently; thus, episodes requiring clinical interventions could be missed. This work investigates the feasibility of measuring bilirubin concentration in whole porcine blood samples using dual-wavelength transmission measurement. A compact and low-cost dual-wavelength transmission measurement setup is developed and optimized to measure whole blood bilirubin concentrations. Using small volumes of whole porcine blood (72 µL), we measured the bilirubin concentration within a range corresponding to healthy individuals and cirrhotic patients (1.2-30 mg/dL). We demonstrate that bilirubin levels can be estimated with a positive correlation (R-square > 0.95) and an accuracy of ±1.7 mg/dL, with higher reliability in cirrhotic bilirubin concentrations (> 4 mg/dL) – critical for high-risk patients. The optical and electronic components utilized are economical and can be readily integrated into a miniature, low-cost, and user-friendly system. This could provide a pathway for point-of-care monitoring of blood bilirubin outside of medical facilities (e.g. patient's home).

## I. INTRODUCTION

LIVER disease, including its most severe form, cirrhosis, is the leading cause of liver-related deaths. Globally over 1.5 million lives are lost annually, making it the eleventh most common cause of death and the third most common in people aged between 45 and 65 [1, 2]. In the initial stages most patients are asymptomatic, and cirrhosis is usually missed and often diagnosed during medical encounters for other reasons. Following diagnosis, patients are required to attend regular outpatient visits for monitoring and prognostic purposes. Clinicians use specialized scores such as the Model for End-stage Liver Disease and the Child-Turcotte-Pugh scores to assess liver disease severity [3]. However, these scoring systems are complex and require specialized laboratory equipment and expertise [4]. While these methods are essential in clinical settings, liver health may deteriorate between appointments; thus, episodes that would require clinical intervention may be missed. Therefore, an effective point-of-care system for liver health monitoring, which can be used by a community health provider or patients in their homes or place of residence, is needed. Such a system would enable frequent assessment of liver health, leading to early diagnosis and improved prognostic outcomes.

Bilirubin is a critical biomarker in the assessment of liver health. Indeed, several studies have suggested that bilirubin can be used as a specific and sole biomarker of hepatic failure and mortality risk in cirrhotic patients [5-7]. Several bilirubinometric techniques, including transcutaneous and conventional laboratory-based optical and chemical methods, have been developed and are currently clinically viable [8, 9]. However, although laboratory-based methods are the gold standards [9], they present several limitations, including their relatively large size and the need for trained personnel to operate the instruments and extract reliable estimation of the bilirubin levels. Additionally, patients must attend a hospital or clinic where blood is drawn for bilirubin measurement. This may be unsuitable or impractical for patients in rural or remote locations with no access to a testing facility. This is especially important in emergencies where a rapid evaluation of results is needed to ensure a patient's survival.

In parallel to laboratory-based methods, the transcutaneous optical bilirubinometric technique is currently in clinical use. This technique exploits the fact that bilirubin exhibits strong optical absorption bands distinct on the blood spectrum. Although this technique is noninvasive, it has been shown to be successful only in neonates but ineffective in adults; this is mainly due to increased age-related skin composition (fatty cells and melanin) and thickness, leading to significant optical losses at the wavelengths of interest [10, 11]. Nevertheless, this paved the way for investigating alternative optical bilirubinometric techniques in the adult population, including exploring new detection methods for cirrhosis patients. This paper investigated a quantitative point-of-care bilirubin monitoring device utilizing dual-wavelength (DWL) measurement in blood. We identified two key wavelengths of 470 and 525 nm for bilirubin measurement based on bilirubin and blood's distinct absorption signatures. We then performed DWL measurements in the absorption mode on both blood phantoms and whole blood to estimate bilirubin levels over the 1.2 - 30 mg/dL concentration range. This range represents the physiological and pathological levels from a healthy individual to a cirrhotic patient [12, 13]. Our findings suggest that bilirubin

This work was supported in part by the Australian Research Council through Linkage Grant LP160101052.

JP Ndabakuranye, A. Ahnood and S. Prawer are with the School of Physics, University of Melbourne, Parkville, VIC 3010, Australia. JP Ndabakuranye, A. Ahnood are also with the School of Engineering, RMIT University, Bundoora, VIC 3083, Australia. A. E. Rajapaksa is with Murdoch Children's Research Institute, Parkville 3052, VIC, Australia, Department of Paediatrics, University of Melbourne, Parkville, VIC 3010, Australia, The Royal Children's Hospital, Parkville, VIC 3052, Australia, The Royal Women's Hospital, Parkville, VIC 3052, Australia and Think Project Global, Clayton, VIC 3168, Australia. G. Burchall is with the School of Health and Biomedical Sciences, RMIT University, Bundoora, VIC 3083, Australia. S. Li is with the Institute of Materials Research and Engineering, A*STAR (Agency for Science, Technology and Research), 138634, Singapore. *Corresponding author: A. Ahnood (email: arman.ahnood@rmit.edu.au).

can be determined with high accuracy using our proposed DWL method; this lays the groundwork for the development of an easy-to-use home testing kit that would work with even a single drop of blood.

## II. EXPERIMENTAL SECTION

### A. Materials

**Bilirubin (BR), hemoglobin (Hb), and Human Serum Albumin (HSA) standards:** Bilirubin ($C_{33}H_{36}N_4O_6$, product number: 14370, grade: purum, purity: ≥95%, MW: 584.67 g·mol$^{-1}$), human hemoglobin (product number: H7379), and albumin (product number: A6612, purity: protein, ≥90% biuret, impurity: ≤6.7 EU/g, Endotoxin; Molecular Mass ≈ 65 kDa) were procured from Sigma-Aldrich (Castle Hill, NSW, Australia). BR was supplied in a brown (amber) vial under reduced pressure to avoid any bilirubin photodegradation. Hb standard was purchased in powder form and was predominantly comprised of methemoglobin (MetHb). HSA came in lyophilized powder form.

**Blood sample:** Porcine blood (drawn via venipuncture of the jugular vein) was procured from Diamond Valley Pork (Laverton North, VIC, Australia) immediately after sacrifice.

**Other chemicals:** Chemicals used in cleaning or solution preparation were of analytical grade. These include sodium hydroxide (NaOH), tris-hydrochloride (Tris-HCl), acetone, isopropyl alcohol (IPA), perchloric acid ($HClO_4$), dimethyl sulfoxide (DMSO), sodium hypochlorite (NaClO), sodium citrate (NaCHO), and milli-Q water (18.2 MΩ·cm).

### B. Stock and working standard solutions preparation

In all cases, solutions were prepared in the open air at room temperature in a room illuminated with a dim red light. Solutions were immediately wrapped in aluminum foil and stored in a refrigerator (0-5° C) to avoid unwanted bilirubin photodegradation due to exposure to the ambient light's blue spectral component. Working solutions were stored in polyethylene containers. Unless stated, the pH was not modified. For consistency in units, mass concentration, mg/dL, was used instead of molar concentration, μmol/L, (1 mg/dL = 17 μM for bilirubin).

*1) Working range of bilirubin concentrations*

The erythrocyte breakdown results in the daily production of 250-350 mg of unconjugated bilirubin (UCB) in healthy adults [12]. Via excretory mechanisms, UCB is modified in the liver to become conjugated bilirubin (CB), which is excreted through the large intestines or kidneys. The body maintains normal bilirubin levels between 0.15 and 1.2 mg/dL but can fluctuate up to 2.95 mg/dL, of which more than 75% is UCB [12]. However, pathological disorders (such as cirrhosis) result in elevated bilirubin levels that can rise to 30 mg/dL or higher [13]. Based on this information, 1.2 - 30 mg/dL was used as the working bilirubin concentration range.

*2) Solvent selection*

Bilirubin standard did not dissolve when directly mixed with porcine blood both with and without albumin (particulate bilirubin was observed in the blood); thus, a solvent was required. Common solvents include NaOH, chloroform and DMSO. The first two were not recommended due to the possibility of globin protein denaturation at elevated pH (NaOH) [14] and the risk of poisonous inhalation and environmental damage [15], and cytotoxicity [16] (chloroform). DMSO was the ideal solvent due to its bio-friendliness [17]. DMSO's non-lipophilic characteristics [18] and the finding by *de Abreu Costa, et al.* [19] that cells' exposure time to DMSO is more harmful than DMSO's concentration imply that it does not induce any instant hemolysis. This was backed by the similarity in the absorption spectra of porcine blood with and without DMSO as shown in Fig. S5.c of supporting information (SI). Besides, the average pKa of UCB in DMSO is 11.1 [20] hence the high solubility estimated at ~580 mg/dL [18].

*3) Preparation of pure bilirubin solutions*

Albumin or DMSO was used as solvents to prepare pure bilirubin stock solutions, usually at 30 mg/dL (513 μM) concentration. Accurately weighed quantities of bilirubin (usually 6 mg) were transferred to and dissolved in 20 ml of the solvent. Working standard solutions were prepared by subsequent dilutions to yield the required concentrations.

*4) Preparation of bilirubin-blood phantoms solutions*

An albumin-bilirubin solution at 3:2 molar ratio ([HSA]= 770 μM and [bilirubin]=513 μM for binding saturation) was prepared by dissolving bilirubin in albumin (usually 6 mg of bilirubin in 20 mL of 770 μM albumin aqueous solution) and followed by subsequent dilutions to yield the concentrations of 30, 20, 15, 10, 5, 2.5, 1.2 mg/dL. It was followed by measuring ~28 mg of Hb and mixing it with 0.2 mL aliquots of BR-HSA solution (=14000 mg/dL) for every dilution (Fig. 1.a). 14000 mg/dL was chosen to mimic the concentration of Hb in healthy adults [21]. However, it was reported that Hb's normal solubility in water is 1 part in 7 [22], which is less than the required concentration, hence stirring was required to increase Hb solubility. Vortex mechanical stirring induced foam formation; this was resolved by mild sonication in a water bath at 30°C.

*5) Blood collection and preparation of bilirubin-blood samples*

In our experiments, porcine blood was used and was preferred as the prominent surrogate for human blood due to its availability, low cost, and biochemical similarity to human blood [23]. It was procured from Diamond Valley Pork abattoir (Laverton North, Victoria, Australia) and was collected in a jug and immediately transferred in a Schott bottle placed in an icebox. The anticoagulant-to-blood ratio of 1:9 (v/v) was used with 4% (w/v) sodium citrate [24, 25]. Samples were then transported to RMIT University (Bundoora, VIC, Australia) and immediately refrigerated (~5°C) to prevent degradation before further analysis. Stock blood samples were stirred for 15 minutes daily to avoid the settling of erythrocytes [26], and samples were always gently shaken for 5 minutes before use. Samples were kept for a maximum of 7 days before being disposed of and replaced with fresh samples to ensure optical stability.

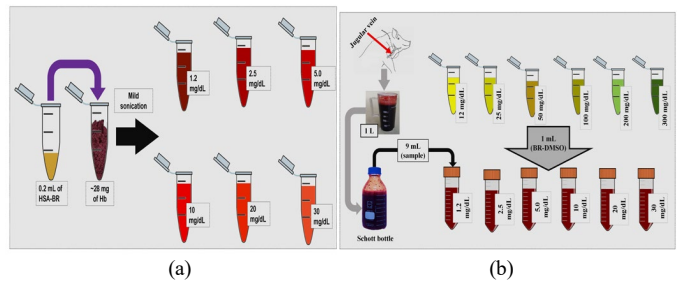

Fig. 1. (a) Steps involved in blood phantoms preparation at varied bilirubin concentrations. (b) Steps involved in collection and preparation of whole blood solutions at various bilirubin concentrations.

Bilirubin-blood solutions were prepared, as shown in Fig. 1.b. Pure bilirubin solutions at 12, 25, 50, 100, 200, and 300 mg/dL concentrations were prepared and transferred to six Eppendorf tubes with concentration labels. Nine (9) mL of anticoagulated blood was poured in each of the six 15-mL polyethylene centrifuge tubes with 1.2, 2.5, 5, 10, 20, and 30 mg/dL labels. A 1-mL volume of the previously prepared bilirubin-DMSO solutions was respectively poured into the 15-mL tubes to yield bilirubin concentrations on the labels. UCB dissolves in DMSO to form a UCB-DMSO solution; hence this process does not involve any changes at the molecular level.

*C. Equipment and measurements*

For bilirubin solutions, measurements were performed using UV-Vis spectrophotometry (300 to 700 nm). For bilirubin-phantoms and bilirubin-blood samples, every sample at a specific bilirubin concentration was measured using both full range spectrophotometry and the DWL approach. Two or three measurements from multiple samples or animal subjects for any given set of conditions were performed to confirm the reproducibility of data, and no variation in measurements was found (coefficient of variation, CV < 2.2 %).

*1) UV-Vis (full range) spectrophotometry*

**Blood phantoms:** Spectral transmission measurements were performed using a UV-Vis spectrophotometer (CARUSO UV-Vis SPECORD 250 PLUS) and a light-field microscope with spectrometer attachment (Nikon Ti-U Microscope combined with IsoPlane 320 Spectrometer). DI water was used for baseline correction. As shown in Fig. S1.a, microfluidic cells were fabricated using 60x22 mm microscope glass coverslips (thickness: 200 µm, ProSciTech) with 0.22-mm thick silicon wafer sheets used as the spacers. LASER cutter (Oxford LASERs Alpha-Series; 532 nm Nd: YAG) was used to cut the silicon wafer and glass coverslips into adequate sizes (usually 22x3 mm for silicon and 22x22 m for glass). The cut pieces were cleaned in acetone, IPA, and water, respectively, for 5 minutes each and dry-cleaned with nitrogen. Two 22x3 mm silicon spacers were sandwiched at both extremes between two glass coverslips (22x22 and 60x22 mm) using an epoxy adhesive (Araldite) and thermally treated at $100^O$ C for 5 minutes. The thickness of microfluidic cells was measured using a digital Vernier-Caliper.

**Whole blood:** UV-Vis Spectrophotometer (Perkin Elmer Lambda 25 Spectrophotometer), O-demountable Quartz microfluidic cell (maximum capacity: 72 uL, type 20, 0.2 mm Light-Path, FireflySci) shown in Fig. S1.b (SI) and a 3-D printed cuvette-like cell holder (Zortrax M2000 plus) were used for full range spectrophotometric measurements on bilirubin-blood samples. DMSO was used as a reference for baseline measurement.

*2) Benchtop system for dual-wavelength measurement*

**Blood phantoms:** A benchtop system for DWL measurement on phantoms was built (Fig. 2.a). The setup consisted of a Red-Green-Blue Light-emitting diode, RGB LED, (L-154A4SUREQBFZGEW, Kingbright) to produce optical power, a sense resistor and reverse-biased silicon photodiode, PD, (BPW21R, Vishay) to monitor the transmitted light intensity, a voltage source (GW INSTEK GPC-1850D) to provide the electrical power to the LEDs as well as a bias to the PD, a digital multimeter (ESCORT-3146A) to measure the voltage drop across the sense resistor, and a camera (Carl Zeiss Tessar HD 1080p Logitech) to record the DMM display values. The LED and photodiode were fixed at 3.0 cm-separation. A 3-D printed optical stage (Cocoon Create modelmaker printer) in conjunction with a homemade microfluidic cell (maximum capacity: 121 µL, area: 482 $mm^2$, pathlength**:** ~250 µm) was used to hold the blood phantoms. Blue (470 nm) and green (525 nm) light from an RGB LED were used. Note that, although these LEDs have maximum optical power outputs at 470 and 525 nm, respectively, they exhibit wide spectral emissions with the full width at half maximum (FWHM) of 28 nm for blue and 30 nm for green. To minimize the measurement errors due to motion artifact, the cell holder was clamped on the retort stand, and both the PD and the LED were glued at fixed equidistant positions on the cell holder. The spectral characteristics of the used colors are shown in Fig. 5.a, and the optical powers were 802 µW (for blue) and 615 µW (for green).

Aliquots of phantoms were aspirated into the syringe and were then perfused into the cell by capillary force. The cell was placed into a 3-D printed cell cover with top and bottom windows cut to ensure the line of sight between the LED and PD and the side windows to insert the cell containing the phantoms to avoid any optical shunting.

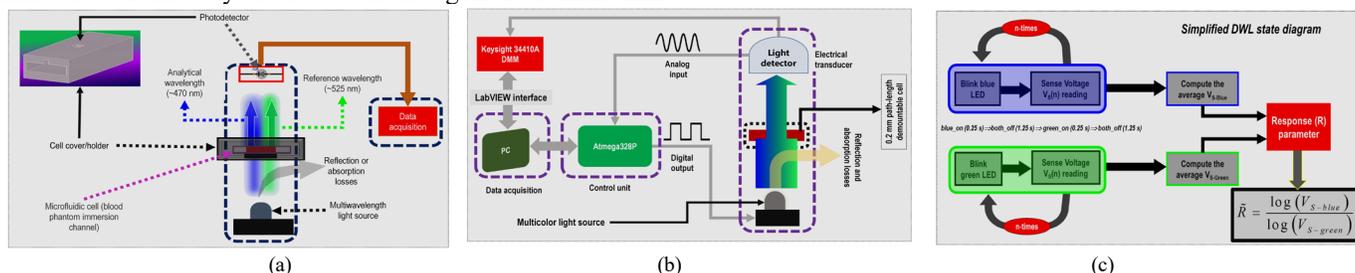

Fig. 2. (a) Illustration of a benchtop system for dual-wavelength measurement on blood phantoms. Manually switched light (470 and 525 nm) was allowed through the blood phantom to the photodiode and converted into a photocurrent measured as a voltage drop across a sense resistor by a bench digital multimeter. (b) Illustration of a benchtop system for dual-wavelength measurement on porcine blood. Pulsed light at 470 and 525 nm was allowed through the blood to the photodiode, converted into an electrical signal, and results were acquired on a PC via a LabVIEW interface (c) State diagram for the dual-wavelength system in (b).

**Whole blood:** A microcontroller (MCU)-based benchtop system for DWL measurement on the whole blood-bilirubin was implemented (Fig. 2.b). It consisted of a MCU (ATmega328P, Microchip), a 200-μm path length microfluidic cell, an RGB LED (L-154A4SUREQBFZGEW, Kingbright), a light sensor (TEMT6000, Vishay), a benchtop 6½ Digit Digital Multimeter (Keysight 34410A) and a voltage/current controlled power supply (Aplab 9 in 1 test lab, model 4049 Variable Voltage, and Current Source). The light source and the detector were placed 5 cm apart, and the irradiance intensity was 992 μW/cm2 (for blue) and 678 μW/cm2 (for green). Light pulse on-times and off-times were selected as 250 and 1250 ms. The number of pulses used for each color was 5 to minimize bilirubin photodegradation while still allowing for an SNR exceeding 16 dB. See fig. S2 a & b (SI).

Aliquots of bilirubin-blood mixtures at the previously specified bilirubin concentration were pipetted and gently added onto the demountable cell. The cell was then transferred onto the 3-D printed cell cover (Zortrax M200 Plus 3D Printer), whose window mimicked the surface outline covered by a blood sample (area: 360 mm2) to avoid optical shunting. The cell and the cell cover were then moved onto the measurement stage between the light source and the detector.

*3) Other equipment*

FieldMax II LASER power meter (UV/Vis/IR, Coherent inc.) for optical power measurements. USB2000+ Spectrometer (250-700 nm, Ocean Optics) with a fiber optic connector for spectral measurement. Milli-Q UV (185 nm) water purification system (8.2 MΩ.cm) as a DI water source. Optical profilometer to measure the cell path length. Vortex mixer and magnetic plate for mixing, ultrasonicator for cleaning and mixing, analytical balance for mass measurement, and pH-meter for pH adjustment.

## III. RESULTS

### A. Feature selection and validation

Several features (Table 1) must be optimally selected to determine the concentration of bilirubin in blood phantoms or whole blood by the DWL technique).

*1) Analytical and reference wavelengths*

A prerequisite for an effective DWL measurement is selecting analytical ($\lambda_1$) and reference ($\lambda_2$) wavelengths at which the absorbance ratio is directly proportional to the concentration of bilirubin and independent of the interfering components over the concentration range of interest. The DWL concept is inherently based on Beer-Lambert's law, which relates the concentration C [mol/L], and the absorbance A of a substance (t: path length [cm], ε: extinction coefficient, T: transmittance, %). See Fig. S3.a & b (SI) and equation (1). The reagent concentration in the mixture can be correlated with the absorbance ratio at $\lambda_1$ and $\lambda_2$. As illustrated in Fig. S3.c & d (SI) [27], these points can be obtained by determining the points of the maximum and negligible or isosbestic absorbance on the absorption spectrum [28, 29].

Bilirubin does not significantly absorb green light but strongly absorbs blue light ($\varepsilon_{525}$ =214, $\varepsilon_{460}$= 53869 [cm$^{-1}$ M$^{-1}$]), as illustrated in Fig. 3.a&b. However, the blue spectrum is also a region of high absorption for Hb, lipids, and melanin [30]. As highlighted in Fig. 3.b, at a 97% oxygen saturation, bilirubin's extinction coefficients are higher than Hb's within the wavelength range from 453 to 482 nm. Indeed, 470 nm is the optimal choice as an analytical wavelength as it provides the largest difference between the extinction coefficients of bilirubin and Hb. Moreover, as illustrated in fig. S4.a (SI), ~525 nm is an isosbestic point for oxyhemoglobin (HbO) and deoxyhemoglobin (HbR). This means that the change in oxygenation level does not affect the absorption at 525 nm. Critically, bilirubin exhibits a weak absorption at 525 nm compared to HbR and HbO, as shown in Fig. 3.a. Thus, $\lambda_1$ and $\lambda_2$ were selected as 470 and 525 nm due to high absorption for bilirubin under blue light, weak absorption for bilirubin under green, the existence of an isosbestic point for HbR and HbO at 525 nm and their off-the-shelf availability.

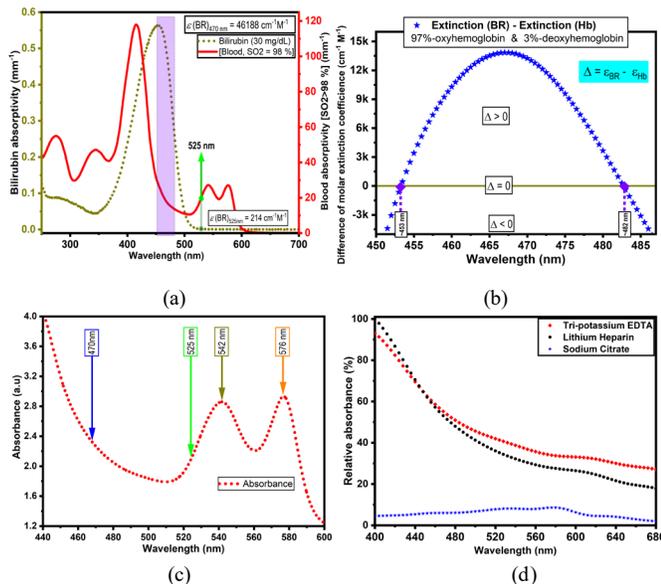

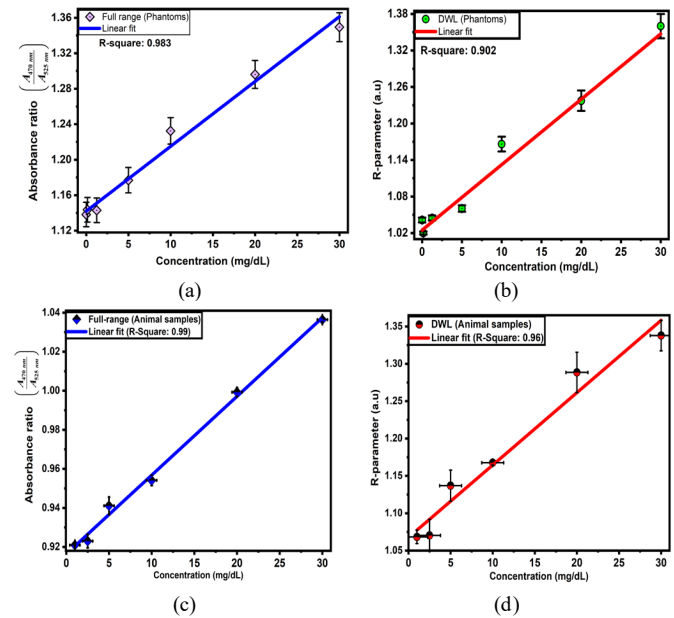

Fig. 4. (a) Comparison of bilirubin and theoretical blood extinction coefficients. Bilirubin has a negligible absorptivity under green and red illumination but highest under blue (b) the difference between bilirubin and hemoglobin molar extinction coefficients. The maximum value of the difference occurs at 470 nm (c) Optical absorption spectrum of whole blood with no added bilirubin shows that 200 μm pathlength is sufficiently thin for the absorption measurement at 470 and 525 nm. The two indicated absorption peaks at 542 and 576 nm correspond to HbO β and α bands, respectively. (d) Illustration of anticoagulant selection. Sodium citrate does not interfere in the visible range.

Fig. 3. Graphs of ratios with respect to bilirubin concentration (a) Absorbance ratio ($A_{470}/A_{525}$) extracted from full-range spectral measurement on phantoms. (b) R-parameters obtained from DWL measurement on phantoms. (c) Absorbance ratio ($A_{470\ nm}/A_{525\ nm}$) extracted from full-range spectra of whole blood. (d) R-parameters obtained from DWL measurements on whole blood.

Given that Hb is the major light-absorbing blood chromophore [30], blood absorbance at 470 nm and 525 nm is shown in equation (2). The absorbance ratio can be correlated with bilirubin concentration. $\varepsilon_{BR, 470}$, $\varepsilon_{BR, 525}$, $\varepsilon_{Hb, 470}$, $\varepsilon_{HB, 525}$ are the molar extinction coefficients for bilirubin or hemoglobin at 470 or 525 nm. $C_{BR}$, $C_{Hb}$: molar concentration for BR or Hb).

$$A = \varepsilon * t * C = 2 - \log_{10}^{T} \Big\|_{T=100*\frac{I}{I_0}} \quad (1)$$

$$\begin{cases} A_{470} = \varepsilon_{BR, 470} * C_{BR} * t + \varepsilon_{Hb, 470} * C_{Hb} * t \\ A_{525} = \varepsilon_{BR, 525} * C_{BR} * t + \varepsilon_{Hb, 525} * C_{Hb} * t \end{cases} \Leftrightarrow R = \frac{A_{470}}{A_{525}} \approx C_{BR} \Big\|_{\varepsilon_{BR, 525} \approx 0} \quad (2)$$

*2) Path length*

A suitable path length must be determined to obtain an optimal transmitted intensity and a useful electrical signal with an adequate signal-to-noise ratio. As a major light-absorbing pigment of the blood, Hb has an impact on the optimal path length. For the purposes of determining this length, we kept the Hb concentration at normal levels in healthy adults (~14000 mg/dL) [21]. Several path lengths were investigated, and results showed that path lengths below 300 µm are suitable for transmittance measurements. Longer path lengths resulted in too much absorption and hence in higher spectral noise. Further information along with results are provided in Fig. S4.b&c (SI). As illustrated in Fig. 3.c, a useful signal at both 470 and 525 nm was obtained for raw blood absorption measurements at 200 µm path length.

*3) Animal blood sample validation*

The porcine blood was selected based on its biochemical similarity to human blood [23]. Optical validation of procured whole blood was required and involved locating the absorption peaks and ascertaining the oxygen saturation (SaO$_2$) levels. As shown in Fig. 3.c, a 200 µm-path length spectrophotometric measurement showed two clear peaks at 542 and 576 nm wavelengths corresponding to β and α HbO absorption bands, respectively. Moreover, a regression analysis showed that the SaO$_2$ of used blood was 68% (R-Square: 0.92), which agrees with the venous SaO$_2$ in healthy pigs, which is generally between 55% and 75% [31].

An anticoagulant was required to stop blood clotting and optical degradation. Lithium heparin (LH), tri-potassium ethylene diamine tetra-acetic acid (K3-EDTA), and sodium citrate (NaCHO) were considered as potential anticoagulants. Nonetheless, LH and K3-EDTA were ruled out due to their possible optical interference with bilirubin spectra in the visible spectrum as shown in Fig. 3.d; thus, NaCHO was preferred and utilized as an anticoagulant for our application. Experiments also showed that blood was more optically stable with an anticoagulant. Blood sedimentation was observed in the absence of anticoagulant and resulted in the absorption peaks reduction, fig. S5 (SI).

TABLE 1
OPTIMAL FEATURES SELECTED FOR DWL MEASUREMENT

| Method parameter | Optimized parameter |
|---|---|
| Operating wavelengths | 470 and 525 nm |
| Pathlength | 200 µm |
| Solvent | Albumin and DMSO |
| Bilirubin concentration range | 1.2 – 30 mg/dL |

*B. Linearity and range: bilirubin and Beer's law*

The adherence of bilirubin to Beer's law was investigated over the concentration range of 1.2-30 mg/dL. Aliquots of appropriate bilirubin dilutions were scanned on a light-field spectrometer. The absorbance at 470 and 525 nm was extracted from the overlain spectra and recorded as the net bilirubin absorbance. The standard curves for both 470 and 525 nm are plotted, and these results ascertain a strong positive linear correlation between the absorbance of bilirubin and its concentrations. The standard calibration curves and overlain spectra are shown in Fig. S6 (SI).

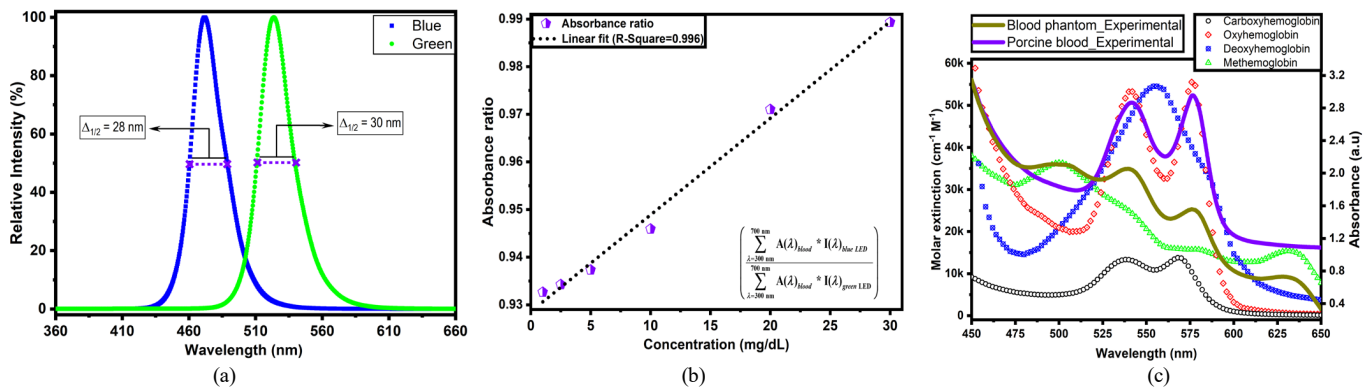

Fig. 5. (a) Relative spectral intensity I(λ) distribution of used LEDs. The measurement was done at forward current IF = 10 mA. (b) The potential effect of LED's spectral variability on bilirubin sensitivity in blood. This was achieved by integrating the absorbance spectra of blood over blue and green LEDs' spectral power and computing the absorbance ratio. (c) Comparison between experimental blood phantoms and porcine blood spectra with the literature data for Hb forms and blood. Regression analysis shows that blood phantoms are made up of 75% MetHb, 22% HbO, 3% HbR, and 0.0% COHb whereas porcine blood consisted of 68% oxygenated blood and 32% deoxygenated blood. MetHb: Methemoglobin; HbO: Oxyhemoglobin, HbR: Deoxyhemoglobin; COHb: Carboxyhemoglobin.

### C. The sensitivity of bilirubin in blood phantoms

Bilirubin sensitivity in phantoms was investigated using full range spectrophotometry and DWL (blue and green) measurements. Full range measurements were performed using a light-field spectrometer. The absorbance values at wavelengths 470 and 525 nm were then extracted from the overlain spectra and used to calculate the absolute absorbance ratios ($A_{470\ nm}/A_{525\ nm}$). Plots are shown in Fig. 4.a. DWL measurements were performed on a benchtop system shown in Fig. 2.a. Sense voltages for blue and green were measured, and the ratio of their logarithms is computed as shown in equation (3) and captured by what we denote as R-parameter $\tilde{R}$ in this paper. Plots of $\tilde{R}$ vs [BR] are shown in Fig. 4.b.

In Fig. 4.a & b, ratios were plotted as a function of bilirubin concentration, and regression analysis shows a very high linear correlation and even stronger at higher bilirubin concentrations (R-square > 0.99). The coefficients of determination for both methods are summarized in Table S1.

### D. The sensitivity of bilirubin in porcine blood

The sensitivity of bilirubin in porcine blood was investigated using both full range spectrophotometry and DWL measurement. Full range measurements were performed using a UV-Vis spectrophotometer. The extracted absorbance at 470 and 525 nm were used to compute the absolute absorbance ratios ($A_{470\ nm}/A_{525\ nm}$) and were plotted as a function of bilirubin concentration, as shown in Fig. 4.c. The DWL system in Fig. 2.b was used to measure the sense voltages. These voltages were used to compute R-parameters using equation (3) and were plotted as a function of bilirubin concentrations in Fig. 4.d. In Fig. 4.c & d, results showed a linear relationship with coefficients of determination greater than 0.95. These are summarized in Table S1 (SI).

$$\tilde{R} = \frac{\log\left[V_{sense(blue)}\right]}{\log\left[V_{sense(green)}\right]} \quad (3)$$

## IV. DISCUSSIONS

### A. Uncertainty and error analysis

#### 1) Mass and path length measurement

The sensitivity studies on phantoms involved more than 50 samples 28.0±0.8 mg each, with 250 ±12 µm path lengths. Equation (1) shows the relationship between the absorbance, path length, and mass. Since the volume was maintained at 0.2 mL, the source of error was due to mass and path length. The overall absorbance error was estimated at ~1.4%.

#### 2) Light emitting diodes' spectral widths

The LEDs used were not ideal monochromatic sources as they exhibit a wider spectral range of 28 and 30 nm width at 50% of the maximum, respectively, for blue and green LEDs (Fig. 5.a). The effect of spectral variability on bilirubin sensitivity was studied by calculating the impact of an LED spectrum on the absorbance ratio. We integrated the blood sample absorbance spectra, obtained by spectrophotometer, over the wavelengths occupied by the LED's relative spectral power, as shown in equation (4). As shown in Fig. 5.b, the slope of the line of fit was 0.002 per mg/dL, which is comparatively the same as that in Fig. 4.c & d. Hence the use of an LED does not induce significant errors.

$$\frac{\sum_{\lambda=300\ nm}^{700\ nm} A(\lambda)_{blood} * I(\lambda)_{blue\ LED}}{\sum_{\lambda=300\ nm}^{700\ nm} A(\lambda)_{blood} * I(\lambda)_{green\ LED}} \quad (4)$$

Besides HbO and HbR, carboxyhemoglobin (COHb) or methemoglobin (MetHb) may exist in blood at low concentrations, owing to their unique absorption signatures that may cause false interpretation of bilirubin concentration. A regression analysis of phantoms' spectra was performed, and results show that MetHb was most predominant, as evidenced by the comparison between the theoretical [32] and experimental data as shown in Fig. 5.c. On the other hand, a regression analysis of spectral data showed that animal samples consisted of HbO at 68% and HbR at 32% (Fig. 5.d). Other parameters such as oxygen saturation and hematocrit and their effect on bilirubin concentration quantification were not investigated in this study.

TABLE 2
COMPARISON OF THE WORK HEREIN TO THE COMPETING TECHNOLOGIES. (TC: TRANSCUTANEOUS, FOC: FIBER OPTIC CABLE, CSF: CEREBROSPINAL FLUID)

| Reference | Sample type | Method | Mode | Volume [µL] | Measured [BR] range [mg/dL] |
|---|---|---|---|---|---|
| This study | Whole blood | 2 wavelengths (470 and 525 nm LEDs) | Transmittance | 72 | 1.2 - 30 |
| Keahey [34] | Plasma | 3 Wavelengths (470, 590, 660 nm LEDs) | Reflectance | 20 - 80 | 1.1 – 23.0 |
| Jacques [35] | Skin/TC | White light source (mathematical correction at 480 nm) | Reflectance via FOC | Not applied | 1 – 17 |
| Yamanouchi [8] | Skin/TC | White light (Optical filters with 460 & 550 nm cutoffs are used) | Reflectance via a FOC | Not applied | 2.5 – 19.0 |
| Smith [36] | CSF | Incandescent light (scanned and extracted at 340, 415, & 460 nm) | Spectroscopic analysis | > 100 | 0.02 – 1.8 |
| Zabetta [37] | Plasma | 2 wavelengths (blue and green LEDs) | Reflectance | 25 | 8.17 – 18.25 |

*3) Porcine blood residual bilirubin*

Experiments were performed on blood phantoms and whole blood, and the agreement between both suggests that the residual bilirubin in porcine blood has a minimal effect. Besides, as mentioned, the normal bilirubin is in the [0.1 - 1 mg/dL] range [12]; hence the residual bilirubin in porcine blood is expected in that range. This consistent residual bilirubin presents a systematic error and, if required, can be accounted for in the postprocessing steps.

*4) Porcine blood residual bilirubin*

Experiments were performed on blood phantoms and whole blood, and the agreement between both suggests that the residual bilirubin in porcine blood has a minimal effect. Besides, as mentioned, the normal bilirubin is in the [0.1 - 1 mg/dL] range [12]; hence the residual bilirubin in porcine blood is expected in that range. This consistent residual bilirubin presents a systematic error and, if required, can be accounted for in the postprocessing steps.

*5) Hemoglobin levels variability*

Blood Hb concentration can vary for about 4.5 g/L (i.e., 140 ± 4.5 g/L) on average in adults [33]. To quantify the effects of Hb levels variations on the bilirubin measurement, the absorbance spectrum of whole blood with varying Hb concentration (135 to 145 g/L) was analyzed. The effect of this variation on the extracted bilirubin levels within the range of 5 mg/dL to 30 mg/dL was assessed – see Fig S7 a, b, and c. As illustrated in Fig S7 d, this variation results in a small uncertainty in the detected bilirubin concentration. The error is largest at higher bilirubin levels with a value of ±0.99 mg/dL.

*B. Analytical correlations between dual-wavelength and full range spectrophotometry*

Fig. 4.a-d show the linear correlation between the absorbance ratio or R-parameters and bilirubin concentration on the entire bilirubin concentration range. Comparing both methods on both phantoms and whole blood shows a similar correlation. Although results were in tight correlation and compared well, full-range scanning exhibited a slightly higher coefficient of determination than DWL.

*C. Comparison with competing technologies*

Table 2 provides a comparison between this work and several previously reported *ex-vivo* studies that use optical methods for bilirubin quantification. The comparison is based on parameters including sample type, technology (method and mode), sample volume and the measured bilirubin concentration range. In contrast to the comparative studies, our study uses whole blood samples, and the measurement is performed in the transmittance mode. Besides, the bilirubin concentration range reported herein was wider than those in the competing studies [8, 34-37].

## V. CONCLUSIONS

Diagnosis and prognosis of cirrhosis involve a quantitative measurement of bilirubin, a specific and major biomarker of liver health status. In clinical settings, this requires specialized instruments, reagents, and personnel; however, our study suggests that a DWL method could be used to quantify bilirubin without the limitations posed by current laboratory methods. The feasibility of this technique was investigated by exploring bilirubin sensitivity in phantoms and whole blood samples. Our DWL system showed an overall accuracy of ±1.7 mg/dL on the [1.2-30] bilirubin concentration range. These results lay the groundwork for developing portable point-of-need device/s capable of quantitative (estimating bilirubin levels) or qualitative (distinguishing between normal, elevated, or critically elevated bilirubin levels) measurements with less than a single drop of blood. Considering the size of illumination (~3 mm radius) and optical detection [~1.3 mm radius (TEMT6000)] spots used in this work, a device utilizing a 10x10 mm microfluidic cell with 0.2 mm path length would be practical. This corresponds to a blood volume of 20 µL, which is smaller than a drop of blood (~50 µL). Furthermore, the technique could be optimized to provide additional information, including the temporal (hourly or daily) evolution of bilirubin, at a low cost with simplicity and rapid turn-around-times. Research that validates the DWL technique on human subjects could also revolutionize the disease management strategies and improve prognostic outcomes. This is especially true for those in remote areas or developing countries where access to healthcare facilities is limited.

ACKNOWLEDGMENT

Jean Pierre gratefully acknowledges the Melbourne Research Scholarship, RMIT University (School of Health and Biomedical Sciences and School of Engineering), and Diamond Valley Pork Pty Ltd. This work was performed in part at the Melbourne Centre for Nanofabrication (MCN) in the Victorian node of the Australian National Fabrication Facility (ANFF). The Australian Research Council also supported this work through Linkage Grant LP160101052. Dr. Rajapaksa acknowledges NHMRC's early career fellowship (GNT 1123030). The authors gratefully acknowledge insightful discussions with Dr. Terence Leung.

COMPETING INTERESTS

Authors declare that they have no conflicting financial or non-financial interests.